\documentclass[abstract=on]{scrartcl}

\usepackage[bindingoffset=0mm,left=25mm,right=25mm,top=25mm,bottom=25mm,footskip=14mm]{geometry}
\usepackage[dvipsnames]{xcolor}
\usepackage{titling}
\usepackage{amsmath}
\usepackage{amssymb}
\usepackage[utf8]{inputenc}
\usepackage{url}
\usepackage{authblk}
\usepackage{enumitem}
\usepackage{graphicx}
\usepackage{listings}

\usepackage{caption}
\usepackage[nameinlink]{cleveref}
\crefname{lstlisting}{listing}{listings}
\crefname{algocf}{algorithm}{algorithms}

\usepackage{tikz}
\usetikzlibrary{calc, patterns}
\usepackage{pgfplots}

\usepackage{bm}

\definecolor{NVGreen}{rgb}{0.43, 0.69, 0.26} %
\definecolor{NVRed}{rgb}{0.945, 0.349, 0.373}
\definecolor{NVBlue}{rgb}{0.349, 0.604, 0.827}
\definecolor{NVOrange}{rgb}{0.976, 0.651, 0.353}
\definecolor{NVPurple}{rgb}{.62, 0.4, 0.671}
\definecolor{NVBrown}{rgb}{.804, 0.439, 0.345}
\definecolor{NVPink}{rgb}{.843, 0.498, 0.702}
\definecolor{NVLime}{rgb}{0.745, 0.769, 0.349}

\definecolor{HighlightColor}{rgb}{0.43, 0.69, 0.26}
\definecolor{HighlightColor2}{rgb}{0.945, 0.349, 0.373}
\definecolor{HighlightColor3}{rgb}{0.349, 0.604, 0.827}
\definecolor{HighlightColor4}{rgb}{0.976, 0.651, 0.353}
\definecolor{HighlightColor5}{rgb}{.62, 0.4, 0.671}
\definecolor{HighlightColor6}{rgb}{.804, 0.439, 0.345}
\definecolor{HighlightColor7}{rgb}{.843, 0.498, 0.702}
\definecolor{HighlightColor8}{rgb}{0.745, 0.769, 0.349}
\definecolor{CodeBG}{rgb}{0.95,0.95,0.95}

\usepackage{listings}
\usepackage[ruled]{algorithm2e}
\newlength{\maxwidth}
\newcommand{\algalign}[2]%
{\makebox[\maxwidth][r]{$#1{}$}${}#2$}

\usepackage[detect-weight=true, detect-family=true]{siunitx}
\DeclareSIPrefix{\million}{\text{M}}{2}

\begin{document}

\title{Massively Parallel Stackless Ray Tracing of Catmull-Clark Subdivision Surfaces}%
\author{Nikolaus Binder, Alexander Keller\\NVIDIA}
\date{}

\maketitle

\begin{figure}[!ht]
	\centering
	\vspace*{-2em}
	\includegraphics[height=.3\textwidth, trim = 80 80 190 30, clip]{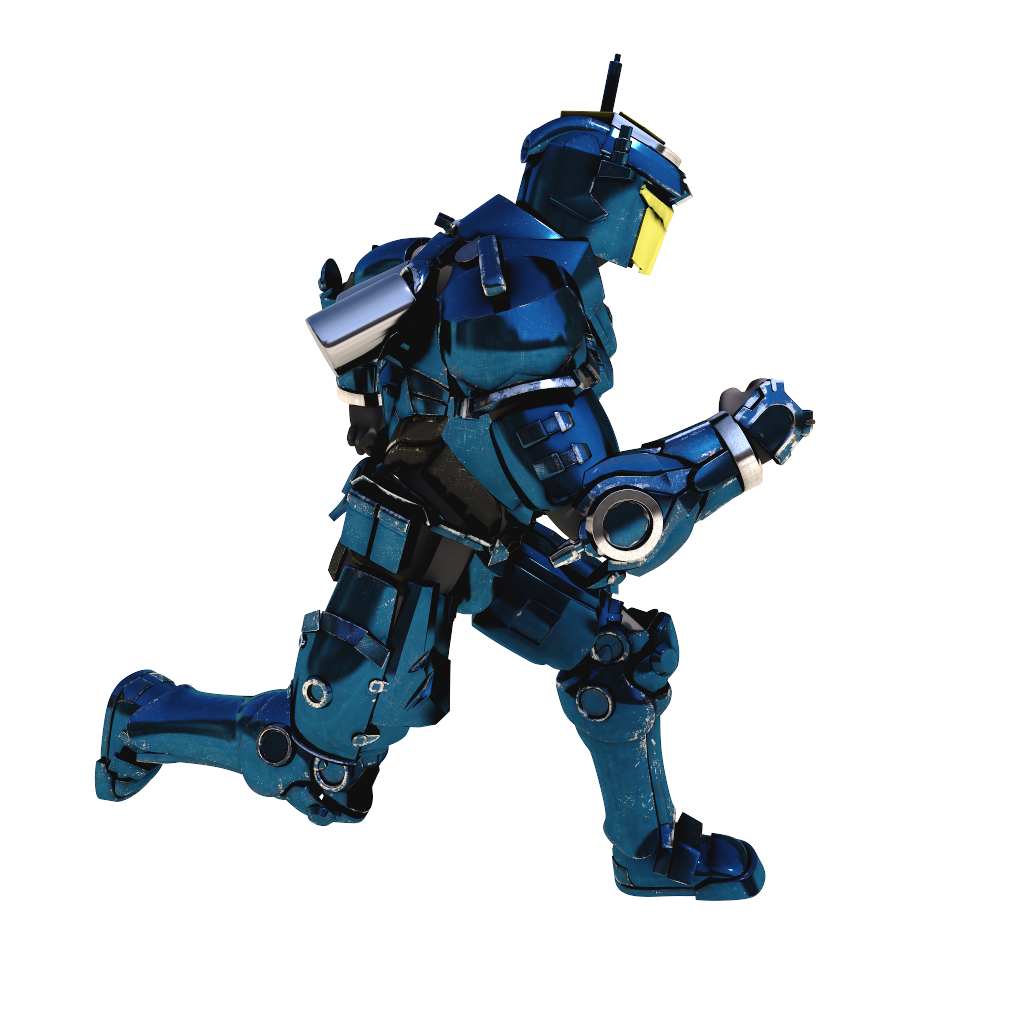} \hfill
	\includegraphics[height=.3\textwidth, trim = 150 30 170 80, clip]{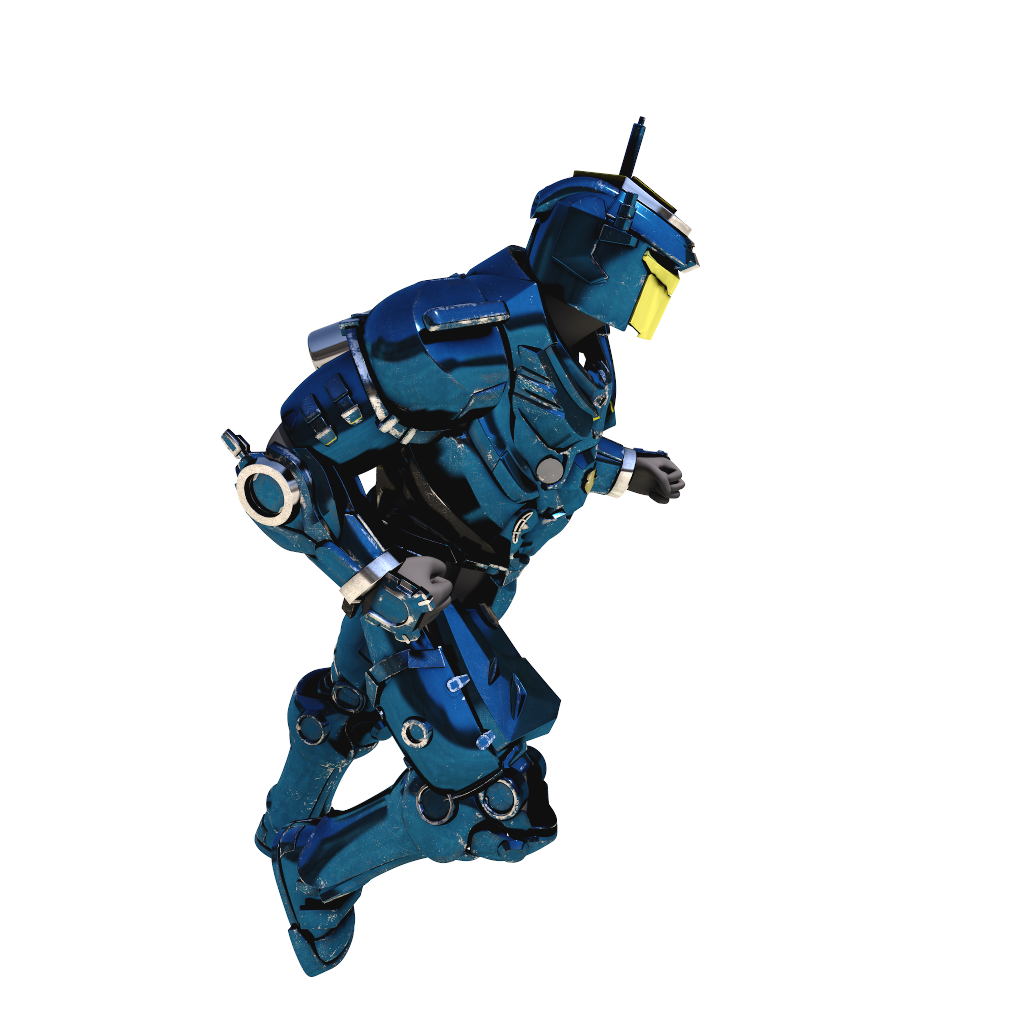}\hfill
	\includegraphics[height=.3\textwidth, trim = 150 30 150 20, clip]{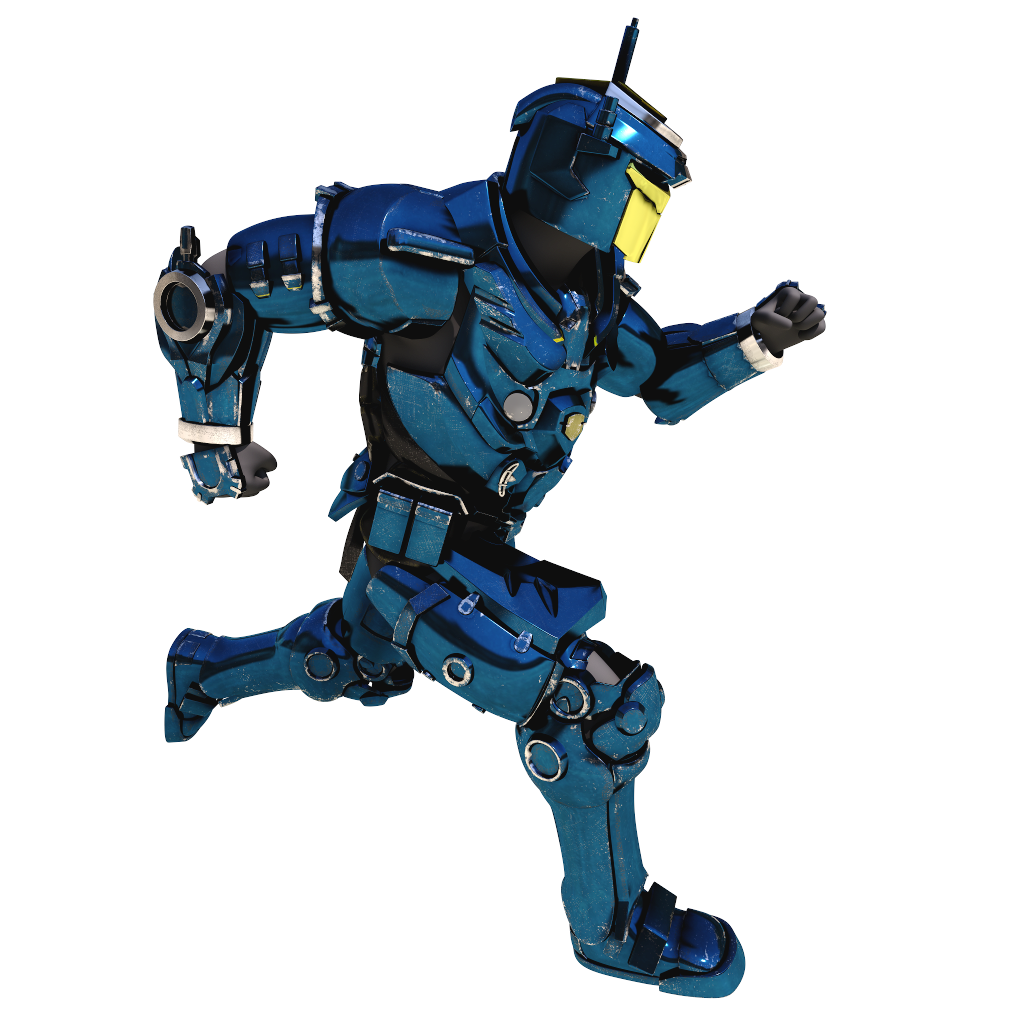} \hfill
	\includegraphics[height=.3\textwidth, trim = 150 30 280 45, clip]{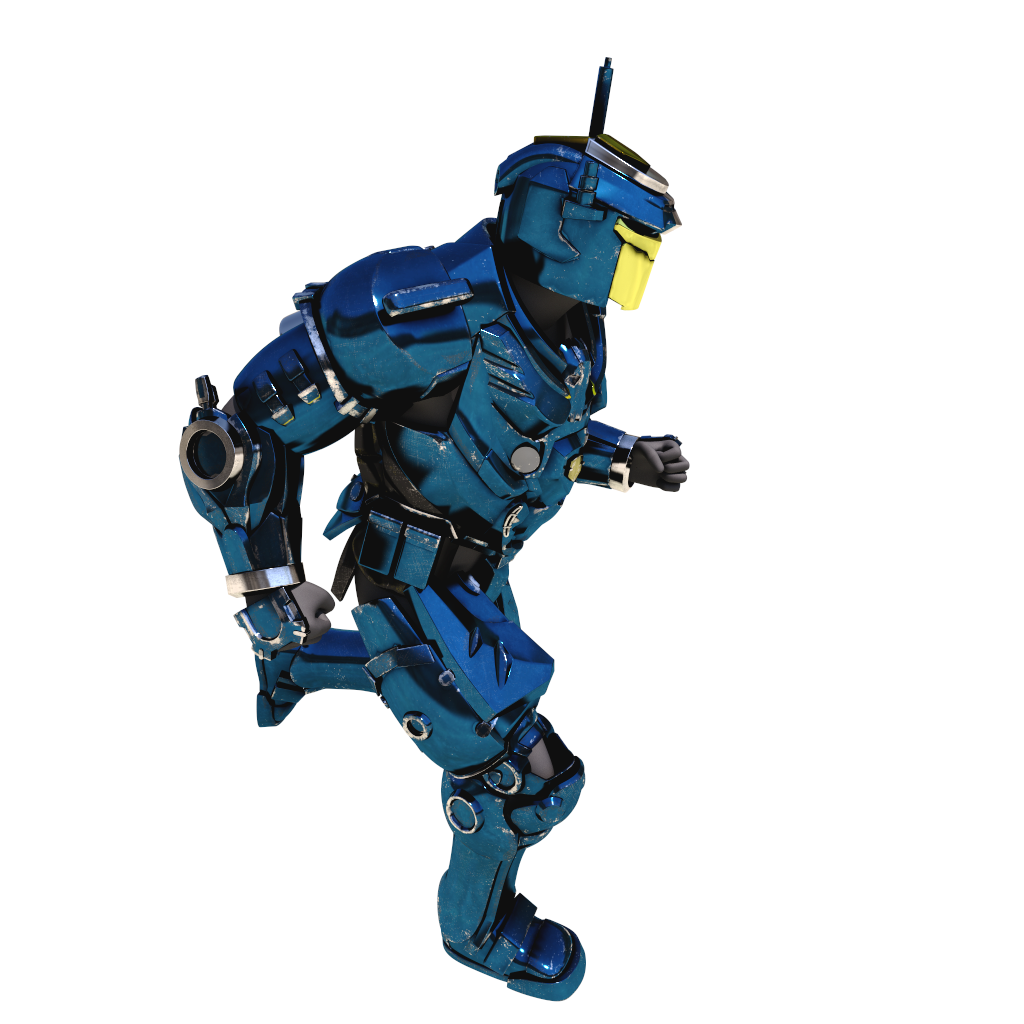}
	\caption{Model composed of Catmull-Clark subdivision surfaces, ray traced with the described direct patch intersection
	method. Performance: $\sim$220 MRays/s for primary and $\sim$85 MRays/s for diffuse/reflection rays on an NVIDIA Titan
	V\texttrademark\ GPU. Armor Guy courtesy of \textcopyright\ 2014 DigitalFish, Inc. All rights reserved.}
	\label{fig:teaser}
\end{figure}

\begin{abstract}
We present a fast and efficient method for intersecting rays with
Catmull-Clark subdivision surfaces. It takes
advantage of the approximation democratized by OpenSubdiv, in
which regular patches are represented by tensor product B\'ezier surfaces
and irregular ones are approximated using Gregory patches. Our algorithm
operates solely on the original patch data and
can process both patch types simultaneously with only a small amount of
control flow divergence. Besides introducing an
optimized method to determine axis aligned bounding boxes of Gregory
patches restricted in the parametric domain,
several techniques are introduced that accelerate the recursive subdivision
process including stackless operation, efficient
work distribution, and control flow optimizations.
The algorithm is especially useful for quick turnarounds
during patch editing and animation playback.
\end{abstract}

\section{Introduction}

For many years, Catmull-Clark subdivision surfaces \cite{Catmull:1978} have been very popular in feature animation. Besides
simple and intuitive modeling, subdivision surfaces facilitate animation because only their control points need to be
animated.
The original construction has been complemented by many extensions, such as semi-sharp creases \cite{DeRose:1998} and
hierarchically defined detail (multi-resolution surfaces) \cite{Pulli:1997}. Over the recent years OpenSubdiv~\cite{OSD}
has been established as the de-facto industry standard for the representation of subdivision surfaces.

While subdivision and evaluation of the parametric surfaces for tessellation of meshes composed from Catmull-Clark
subdivision surfaces in a preprocess is well understood and almost trivially integrates into existing ray tracing
systems supporting triangle meshes, the incoherent memory access patterns of path tracing and memory constraints on
modern graphics hardware are longing for the direct intersection of rays with these surfaces.

\section{Survey of Previous Work}
Rendering systems only computing primary visibility, such as REYES \cite{Cook:1987} or rasterization, rely on
temporary tessellation of subdivision surfaces. Besides using efficient data structures \cite{Pulli:1996} or methods to
determine tessellated vertices without subdivision \cite{Stam:1998}, the efficiency of this procedure stems from the
nature of these systems, which aim to access each surface at most once. In ray tracing based simulations, on the other
hand, access is incoherent, and surfaces may be subsequently hit by different generations of rays. While coherency
within one generation of rays can be improved by sorting \cite{Hanika:2010}, the problem of repeated access across
generations remains. Caching the tessellated data allows for a reuse up to a certain degree \cite{Benthin:2015}, however
the efficiency highly depends on scene complexity, tessellation rate, and cache size. The memory footprint of
hierarchies referencing tessellated primitives can be compressed and compactly represented without pointers due to their
regular quad-tree structure \cite{Selgrad:2016}.

In order to display subdivision surfaces at interactive rates, feature adaptive subdivision \cite{Niessner:2012}
determines a set of patches that is submitted to the hardware tessellation units for rasterization. These patches can
be partitioned into two classes, regular and irregular patches. Patches are regular unless one of their control points
has a valence which is different from four. The limit surface of regular patches can be exactly and efficiently
evaluated using cubic tensor product B-Spline or equivalent B\'ezier surfaces. For the fast evaluation of the limit
surface of irregular patches, however, an approximation using Gregory patches \cite{Gregory:1974} is used. As each
irregular patch with $m$ vertices with a valence different from four and $n$ vertices in total can be subdivided into $n
- m$ regular patches and $m$ irregular ones, irregular patches are subdivided several times in order to decrease the
area in which the approximation is used. A similar refinement is executed for patches next to creases and semi-sharp
edges. When tessellating patches, special care has to be taken whenever the subdivision level or tessellation factor
differs across edges in order to guarantee crack free surfaces.

Methods that directly determine intersections with subdivision surfaces lend themselves to ray tracing simulations. They
neither require parameter tweaking, nor preprocessing, nor additional memory for tessellated geometry and acceleration
structures for this micro geometry.
Solutions for the equality of the parametric representation of a ray and those of a surface patch can be computed using
root finding of the resulting polynomials using Laguerre's method \cite{Ralston:1965,Kajiya:1982} or Newton-Raphson
solvers \cite{Parker:1999}. However, methods based on Newton iterations need a good starting point in order to be
efficient and to guarantee correct results. While for primary rays these starting points can often be guessed from
adjoining rays \cite{Joy:1986}, and regions in parameter space with guaranteed convergence of the solver can be
identified \cite{Toth:1985}, coherence may lead to incorrect closest intersection points \cite{Lischinski:1990}, only
supporting coherent rays is not sufficient, and the cost of identifying safe regions per ray can be prohibitively high
in practice. The convergence of Newton iterations can be accelerated by using polynomial extrapolation and by determining
better starting points from the bounding volume of the patch defined as a trapezoidal prism \cite{Qin:1997}, by adaptively
subdividing the patch into parts enclosed in parallelepipeds and referenced by a hierarchy for pruning \cite{Barth:1993},
or by simply pre-subdividing patches into smaller ones referenced in the same top-level acceleration data structure
\cite{Nigam:2012}. Besides improving performance, these methods also improve the quality of starting points, but either
increase the memory footprint for the extension of the acceleration structure or add a certain performance overhead.

Recursive subdivision of the patch \cite{Kajiya:1983,Whitted:1980} does not suffer from issues of methods based on root
finding. Kobbelt et al. develop a fast on-the-fly subdivision method based on the precomputation of basis functions for
the influence of vertices \cite{Kobbelt:1998}. This, however, needs to be done for all occurring valences and crease
values, resulting in a combinatorial explosion. M\"uller et al. perform adaptive subdivision of the surfaces based on
several criteria \cite{Mueller:2003}. Still, the process is elaborate. Testing the projections of the convex hull of the
1-neighborhood can be efficiently performed in a ray centric coordinate system \cite{Woodward:1989}, however the
resulting bounding volumes are not especially tight. Oriented slabs allow for more accurate pruning of patches than axis
aligned bounding boxes, especially for rather flat ones \cite{Yen:1991}. Depending on the type of patch and subdivision
rules, the effort can be rather low, but may potentially perform redundant computations. Therefore, attempts were made
to amortize this cost across groups of rays either by exploiting coherence of primary rays \cite{Lischinski:1990}, by
predefining bundles of primary rays \cite{Benthin:2007}, or by sorting batches of rays \cite{Hanika:2010}. As the
potential re-use for large scenes, high subdivision depths, and small cache sizes may be small and at the same time the
performance of modern massively parallel processors is very fast compared to memory access, subdividing independently
for each ray individually is getting more and more interesting. Instead of intersecting with micro geometry, nested
hierarchies of axis-aligned bounding boxes of sub-patches determined on demand during recursive subdivision can be
refined until the intersected bounding box is sufficiently small to be identified as point of intersection along the ray
\cite{Pulleyblank:1987,Dammertz:2006}. The absence of an approximation with false negatives after termination of the
subdivision process guarantees watertightness without further effort.

B\'ezier clipping \cite{Nishita:1990} iteratively reduces the parametric interval in which an intersection can be found
by transforming the ray into the parametric $(u, v)$ space of a patch and cropping it to the bounds of the ray in this
space. The original algorithm suffers from several deficiencies that can be resolved \cite{Campagna:1997,Efremov:2005}.
It can also be used in conjunction with Newton-Raphson solvers \cite{Wang:2001}. Tejima et al. combine B\'ezier clipping
with a bi-linear approximation that is used after subdivision termination \cite{Tejima:2015}. While this approximation
reduces the number of subdivision steps in flat regions, the approximation with bi-linear surfaces cannot guarantee
watertight surfaces: Adjacency on the surface does not necessarily imply shared vertices of the bi-linear
approximations. Furthermore, efficient B\'ezier clipping requires rotations that not only increase the computational
cost compared to simple subdivision by bisection, but may also introduce numerical errors visible as cracks in the
surface.
While consuming patches from an OpenSubdiv preprocess generating B\'ezier and Gregory patches, their method can only
handle tensor product B\'ezier patches and therefore needs to approximate the approximation with Gregory patches even
more. While certain parts are optimized for watertightness, the overall method does not guarantee watertightness: As
the final intersection is performed with a bi-linear patch and domain cropping may terminate for different domain sizes,
adjacent bi-linear patches do not necessarily meet at connecting edges. Furthermore, precision issues in B\'ezier
clipping may also introduce cracks since the resulting intervals may also suffer from the imprecision due to the initial
rotation and all subsequent floating point operations. While the required number of iterations is halved compared to
midpoint subdivision, the required effort is significantly higher, especially since midpoint subdivision is extremely
cheap and already reduces the parameter interval exponentially.

We also investigate intersecting rays with the bicubic patches generated by OpenSubdiv instead of using them for
tessellation, but support both B\'ezier and Gregory patches and focus on a method suitable for efficient massive
parallelization on GPUs. First, we explain how spatial bounds for both B\'ezier and Gregory patches can be calculated
efficiently for patches cropped to parametric domains. The calculation of spatial bounds for Gregory patches restricted
to parametric domains is based on a variant of the algorithm described by Miura et al. \cite{Miura:1994}, which we
optimize to calculate slightly looser bounds at a reduced cost. Second, these bounds are used for hierarchical culling in
an iterative intersection algorithm. Finally, we provide details how performance of our massively parallel
implementation can be optimized. With this massive parallelization in mind, stack memory consumption of recursive
approaches becomes considerable and may cause bandwidth issues. Therefore, one of the key characteristics of the
proposed algorithm is that it does not require a stack for backtracking during subdivision of the patch. Instead,
remaining domains are indicated in two bit trails \cite{Hughes:2009}, kept in registers. Bit trails have previously also
been used for backtracking during traversal of bounding volume hierarchies by restarting from the root node
\cite{Laine:2010}, by using parent pointers \cite{Afra:2014}, or by using a perfect hash map \cite{Binder:2016}. We then
calculate control points and/or bounding volumes for remaining domains, whose position and size can be extracted from
these bit trails.

\section{Algorithm}
\label{sec:algorithm}

In a pre-process OpenSubdiv performs feature adaptive subdivision on the input mesh to isolate irregular patches and
output patch data. Note that we do not need to differentiate between regular and transition patches since we do not need
to take special care at boundaries between subdivision levels. After determining conservative spatial bounds of the
patches, a hierarchy referencing the patches is built.
A ray is then intersected with the mesh by first traversing this hierarchy, pruning patches that cannot be intersected
by the ray, and identifying patches potentially intersected by the ray.

In order to determine the first intersection of a ray and a bicubic patch, we partition the parametric unit square
recursively, alternating $u$ and $v$ direction. Restricting the bicubic patches to the resulting subdomains allows one
to determine bounding volumes as derived for B\'ezier patches in \Cref{sec:bezier_bounds} and Gregory patches in
\Cref{sec:gregory_bounds}.

Refining the subdomains on demand results in a sequence of nested axis-aligned bounding boxes and in fact intersecting
only bounding boxes is sufficient to compute intersections up to almost floating point precision.
\Cref{sec:iterative_hierarchical_subdivision} details \Cref{alg:hierarchical_subdivision} that only uses bit trails
instead of a stack and recomputes control points for backtracking. The numerical properties of the algorithm are
discussed in \Cref{sec:numerical_robustness_and_watertightness}, and its efficient parallelization is explained in
\Cref{sec:efficient_parallelization}.

\begin{figure}
	\centering
	\includegraphics[width=0.4\linewidth]{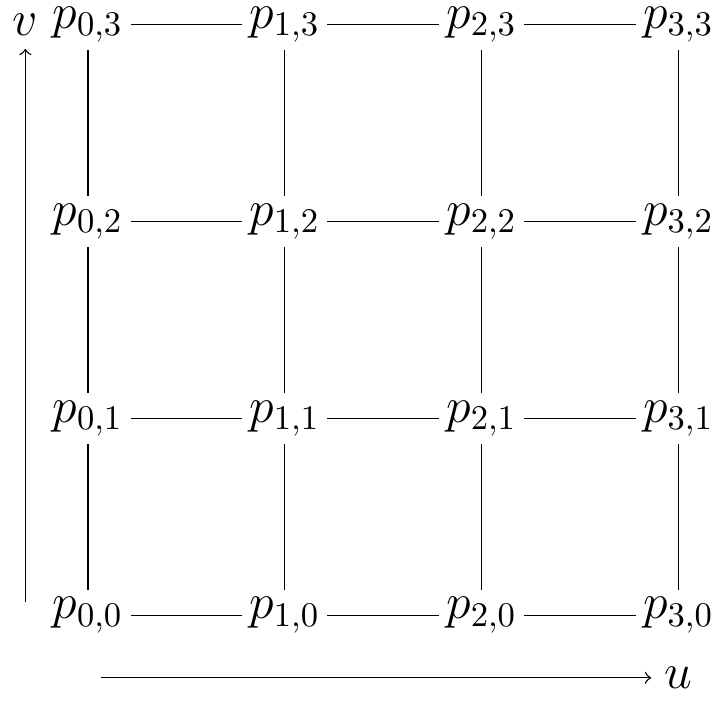}\hfill
	\qquad \includegraphics[width=0.4\linewidth]{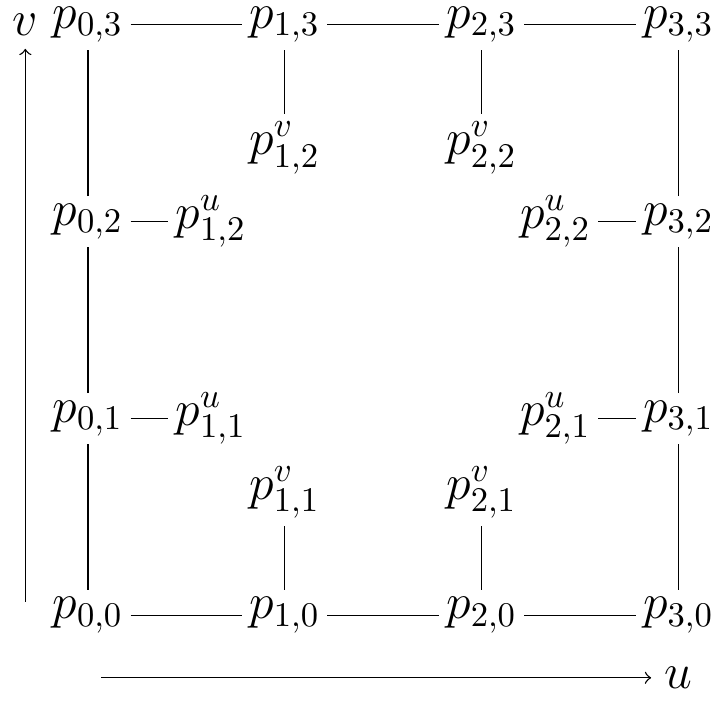}
	\caption{Control vertex layout of a bicubic B\'ezier (left) and Gregory patch (right) along $\bm{u}$ and $\bm{v}$ direction.}
	\label{fig:control_point}
\end{figure}

\subsection{Bounding Bicubic B\'ezier Patches on Subdomains}
\label{sec:bezier_bounds}

A conservative axis aligned bounding box of a bicubic B\'ezier patch is given by the component-wise minimum and maximum of its
control points $p_{i, j}$ $(i,j \in \{0, 1, 2, 3\})$ as guaranteed by the convex hull property of the tensor product surface.

Sub-patches defined by restricting the parametric domain of a B\'ezier patch can also be represented by B\'ezier
patches. Control points of a B\'ezier sub-patch resulting from splitting the domain along a parametric axis can be
calculated using de Casteljau subdivision \cite{deCasteljau:1959}.

A restriction of a bicubic B\'ezier patch $f(u, v)$ defined on the unit square to an arbitrary subdomain $\mathbb{D} :=
[u_0, u_1] \times [v_0, v_1] \in [0,1]^2$ results in a bicubic B\'ezier patch whose control points as laid out in
\Cref{fig:control_point} can be computed by cropping the tensor product surface, for example as implemented in
\Cref{alg:crop_bezier}. Besides intuitively imposing constraints to enforce the properties of the B\'ezier sub-patch,
cropping can be derived by applying de Casteljau subdivision to the patch twice, once to cut each side of one parametric
direction, and repeating the process to crop the other one. In our performance tests, cropping as implemented in
\Cref{alg:crop_bezier} performed slightly faster than the mathematically equivalent successive cropping of parametric
directions.

As splitting a patch using de Causteljau subdivision is significantly faster than computing control poinof a sub
patch from scratch, we only recompute control points after backtracking, i.e. unless the next domain is a subdomain of
the current one.

\subsection{Bounding Bicubic Gregory Patches on Subdomains}
\label{sec:gregory_bounds}

A bicubic Gregory patch is controlled by 20 points as laid out in \Cref{fig:control_point}. Given a pair of
$(u,v)$ coordinates, it can be evaluated by reducing it to a bicubic B\'ezier patch, where the control points on the
boundary are identical and the inner control points
\[
	p_{i, j}(u, v) = g_{i, j}(u, v)p_{i, j}^u + \left(1 - g_{i, j}(u, v)\right)p_{i, j}^v
\]
for $i,j \in \{1, 2\}$ are determined using the weights
\[
	g_{i, j}(u, v) =
	\begin{cases}
		\frac{v}{u+v} & i = 1, j = 1\\
		\frac{v}{(1-u)+v} & i = 2, j = 1\\
		\frac{(1-v)}{u+(1-v)} & i = 1, j = 2\\
		\frac{(1-v)}{(1-u)+(1-v)} & i = 2, j = 2 .
	\end{cases}
\]
Due to the special case $p_{i, j}^u = p_{i, j}^v = p_{i,j}$, a Gregory patch may be considered a generalization of a
B\'ezier patch.

While bounding volumes for Gregory patches can also be determined from the convex hull of its control points,
determining a bounding volume for the restriction to an arbitrary parametric subdomain $\mathbb{D} := [u_0, u_1] \times
[v_0, v_1] \in [0,1]^2$ is not as obvious as in the B\'ezier case, because a sub-patch of a Gregory patch may not
necessarily be representable as a Gregory patch.%

While Gregory patches can be converted to bi-7th degree rational B\'ezier patches \cite{Takamura:1990}, the approach is
ruled out by the additional amount of data, computations, and numerical issues.

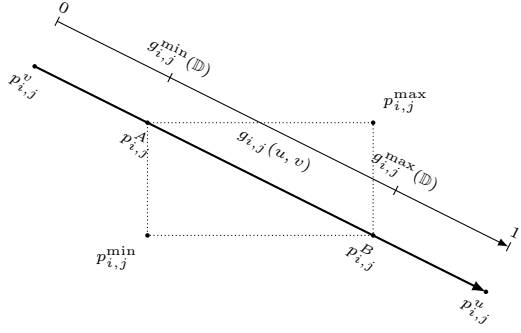
\begin{figure}
	\centering
	\begin{tikzpicture}[scale = 3]
		\fill (0, 1) circle (0.01);
		\fill (2, 0) circle (0.01);
		\draw[-latex, thick] (0, 1) -- (2, 0)
			node[pos=0, font=\tiny, below, sloped] {$p_{i,j}^v$}
			node[pos=1, font=\tiny, below, sloped] {$p_{i,j}^u$}
			node[pos=0.25, font=\tiny, below, sloped] {$p_{i,j}^A$}
			node[pos=0.75, font=\tiny, below, sloped] {$p_{i,j}^B$};
		\draw[-latex] (0+0.1, 1+0.2) -- (2+0.1, 0+0.2)
			node[midway, below, sloped, font=\tiny] {$g_{i, j}(u, v)$}
			node[pos=0.25, above, sloped, font=\tiny] {$g_{i, j}^{\textrm{min}}(\mathbb{D})$}
			node[pos=0.75, above, sloped, font=\tiny] {$g_{i, j}^{\textrm{max}}(\mathbb{D})$}
			node[pos=0, above, sloped, font=\tiny] {$0$}
			node[pos=1, above, sloped, font=\tiny] {$1$};
		\draw (0+0.01+0.1, 1+0.02+0.2) -- (0-0.01+0.1,1-0.02+0.2);
		\draw (0.5+0.01+0.1, 0.75+0.02+0.2) -- (0.5-0.01+0.1,0.75-0.02+0.2);
		\draw (1.5+0.01+0.1, 0.25+0.02+0.2) -- (1.5-0.01+0.1,0.25-0.02+0.2);
		\draw (2+0.01+0.1, 0+0.02+0.2) -- (2-0.01+0.1,0-0.02+0.2);
		\fill (0.5, 0.75) circle (0.01);
		\fill (1.5, 0.25) circle (0.01);
		\draw[densely dotted] (0.5, 0.25) rectangle (1.5, 0.75);
		\fill (0.5, 0.25) circle (0.01);
		\node[font=\tiny, anchor=north east] at (0.5, 0.25) {$p_{i, j}^{\textrm{min}}$};
		\fill (1.5, 0.75) circle (0.01);
		\node[font=\tiny, anchor=south west] at (1.5, 0.75) {$p_{i, j}^{\textrm{max}}$};
	\end{tikzpicture}
	\caption{Bounding the interpolant $\bm{g_{i, j}(u, v)}$ in the domain $\bm{\mathbb{D}}$ also bounds the parametric
	inner control points $\bm{p_{i, j}(u, v)}$.}
	\label{fig:bounding_interpolant}
\end{figure}

However, analyzing the inner control points \cite{Miura:1994} allows for finding bounding volumes of bicubic Gregory
patches restricted to $\mathbb{D}$:
Taking a look at the signs of the derivatives of the weights
\[
\begin{tabular}{l|ll|ll}
$i, j$ & $\frac{\partial}{\partial u}g_{i, j}(u, v)$ && $\frac{\partial}{\partial v}g_{i, j}(u, v)$ &\\[0.5ex]
\hline
$1, 1$ & $-\frac{v}{(u+v)^2}$ & $\leq 0$ & $\frac{u}{(u+v)^2}$ & $\geq 0$\\
$2, 1$ & $\frac{v}{((1-u)+v)^2}$ & $\geq 0$ & $\frac{1-u}{((1-u)+v)^2}$ & $\geq 0$\\
$1, 2$ & $-\frac{(1-v)}{(u+(1-v))^2}$ & $\leq 0$ & $-\frac{u}{(u+(1-v))^2}$ & $\leq 0$\\
$2, 2$ & $\frac{(1-v)}{((u-1)+(v-1))^2}$ & $\geq 0$ & $-\frac{(1-u)}{((u-1)+(v-1))^2}$ & $\leq 0$
\end{tabular}
\vspace*{1em}
\]
allows one to bound the weights
\begin{align*}
	g_{i, j}^\textrm{min}(\mathbb{D}) := \min_{(u,v) \in \mathbb{D}} g_{i, j}(u, v) &=
	\begin{cases}
		\frac{u_1}{u_1 + v_0} & i = 1, j = 1\\
		\frac{(1 - u_0)}{(1 - u_0) + v_0} & i = 2, j = 1\\
		\frac{u_1}{u_1 + (1 - v_1)} & i = 1, j = 2\\
		\frac{(1-u_0)}{(1-u_0) + (1 - v_1)} & i = 2, j = 2
	\end{cases}
\end{align*}
and
\begin{align*}
	g_{i, j}^\textrm{max}(\mathbb{D}) := \max_{(u,v) \in \mathbb{D}} g_{i, j}(u, v) &=
	\begin{cases}
		\frac{u_0}{u_0 + v_1} & i = 1, j = 1\\
		\frac{(1 - u_1)}{(1 - u_1) + v_1} & i = 2, j = 1\\
		\frac{u_0}{u_0 + (1 - v_0)} & i = 1, j = 2\\
		\frac{(1-u_1)}{(1-u_1) + (1 - v_0)} & i = 2, j = 2
	\end{cases}
\end{align*}
on the subdomain $\mathbb{D}$.
Finally, as depicted in \Cref{fig:bounding_interpolant},
\begin{align*}
	p_{i, j}^\textrm{A}(\mathbb{D}) &= 
          g_{i, j}^\textrm{min}(\mathbb{D})p_{i, j}^u 
          + \left(1 - g_{i, j}^\textrm{min}(\mathbb{D})\right)p_{i, j}^v,\\
	p_{i, j}^\textrm{B}(\mathbb{D}) &= 
          g_{i, j}^\textrm{max}(\mathbb{D})p_{i, j}^u 
          + \left(1 - g_{i, j}^\textrm{max}(\mathbb{D})\right)p_{i, j}^v,\\
	p_{i, j}^\textrm{min}(\mathbb{D}) &= 
          \min\{p_{i, j}^\textrm{A}(\mathbb{D}), 
               p_{i, j}^\textrm{B}(\mathbb{D})\}, \text{ and }\\
	p_{i, j}^\textrm{max}(\mathbb{D}) &= 
          \max\{p_{i, j}^\textrm{A}(\mathbb{D}), 
               p_{i, j}^\textrm{B}(\mathbb{D})\} ,
\end{align*}
where minima and maxima are determined component-wise, and hence
\[
p_{i, j}^\textrm{min}(\mathbb{D}) \leq p_{i,j}(u,v)
          \leq p_{i, j}^\textrm{max}(\mathbb{D}) \text{ for } (u,v) \in \mathbb{D} .
\]

As obviously $p_{i,j}^\textrm{min}(\mathbb{D}) = p_{i,j}^\textrm{max}(\mathbb{D}) = p_{i,j}$ for the control points on
the patch boundary, the two bicubic B\'ezier patches defined by the control points on the boundary and $p_{i,
j}^\textrm{min}(\mathbb{D})$ and $p_{i, j}^\textrm{max}(\mathbb{D})$, respectively, bound the volume of the restriction
of a bicubic Gregory patch to $\mathbb{D}$. The bounding volume bounding both bicubic B\'ezier patches restricted to
$\mathbb{D}$ then bounds the bicubic Gregory patch restricted to $\mathbb{D}$.

Instead of maintaining two sets of control points, a conservative bounding volume can be defined by determining the
lower bound from the B\'ezier patch with $p_{i, j}^{\textrm{min}}$ as inner control points and extending its upper bound
by the maximum displacement vector $d_\textrm{max}(\mathbb{D})$, which is defined as the component-wise maximum of the
maximal displacements
\[
  d_{i,j}(\mathbb{D}) := 
  p_{i, j}^\textrm{max}(\mathbb{D}) - p_{i, j}^\textrm{min}(\mathbb{D})
\]
of the inner control points.

Unlike the bounding volumes determined by Miura et al.~\cite{Miura:1994}, we do not use a difference patch based on the
$d_{i, j}$, but instead derive a conservative upper bound of the displacement of the upper bound, which we will
improve in the following:

In floating point arithmetic the displacement is not necessarily monotonously decreasing with increasing subdivision
steps and $\lim_{u_1 \to u_0, v_1 \to v_0}d_\textrm{max}(\mathbb{D}) = 0$ cannot be guaranteed.
This especially happens when $u$ and $v$ are 0 or 1.

The tightness of the bounding volume and its numerical properties can be improved by bounding the tensor product weights
\[
  w_{i,j}(u, v) = B_i^3(u) \cdot B_j^3(v)
\]
in the bicubic B\'ezier surface formula according to $\mathbb{D}$.
For example, analyzing the second Bernstein polynomial of degree 3 yields
\[
\max_{t_0 \leq t \leq t_1} B_1^3(t)
= 
\begin{cases}
	\frac{4}{9} & t_0 \leq \frac{2}{3} \wedge t_1 \geq \frac{2}{3}\\
	3 \cdot t_1^2(1-t_1) & t_1 < \frac{2}{3}\\
	3 \cdot t_0^2(1-t_0) & t_0 > \frac{2}{3}
\end{cases}
\]
when restricted to the interval $[t_0, t_1]$. A similar bound can be derived for $B^3_2(t)$ and hence
\[
  w_{i,j}(u, v) \leq \overline{w}_{i,j}(\mathbb{D}) := \max_{u_0 \leq t \leq u_1} B_i^3(t) \cdot \max_{v_0 \leq t \leq v_1} B_j^3(t)
\]
for $(u, v) \in \mathbb{D}$ and $i,j \in \{1,2\}$. Noting that on the patch boundary $d_{i,j}(\mathbb{D}) = 0$, i.e. for
$i,j \in \{0, 3\}$, the evaluation of a bicubic B\'ezier patch now can be bounded by
\begin{eqnarray*}
\lefteqn{f(u, v) = \sum_{0 \leq i,j \leq 3} w_{i,j}(u, v)p_{i,j}(u,v) } \\ %
	& \leq & \sum_{0 \leq i,j \leq 3} w_{i,j}(u, v)
		p_{i,j}^\textrm{max}(\mathbb{D}) \\
	& = & \sum_{0 \leq i,j \leq 3} w_{i,j}(u, v)
	\left(p_{i,j}^\textrm{min}(\mathbb{D}) + d_{i,j}(\mathbb{D})\right) \\
	& = & \sum_{0 \leq i,j \leq 3} w_{i,j}(u, v)p_{i,j}^\textrm{min}(\mathbb{D}) 
	+ \sum_{i,j \in \{1, 2\}} w_{i,j}(u, v)d_{i,j}(\mathbb{D}) \\
	& \leq & \sum_{0 \leq i,j \leq 3} w_{i,j}(u, v)p_{i,j}^\textrm{min}(\mathbb{D}) 
+ \underbrace{\sum_{i,j \in\{1,2\}} \overline{w}_{i,j}(\mathbb{D})d_{i,j}(\mathbb{D})}_{=:d(\mathbb{D})} .
\end{eqnarray*}
Note that $d(\mathbb{D})$ is now zero in the numerically critical cases ($u_{0/1}$ and $v_{0/1}$ are 0 or 1).

Again, in order to bound a bicubic Gregory patch, we first determine the axis-aligned bounding box of the lower bounding
B\'ezier patch, however, now extending it by displacing its upper bound by $d(\mathbb{D})$, which includes the bounds on
the weights. These computations to determine the control points of the lower bounding B\'ezier patch and the maximum
displacement of the upper bound are subsumed in \Cref{alg:bounding_gregory}.

Now, dealing with both B\'ezier and Gregory patches can be realized by one unified algorithm (see
\Cref{alg:bounding_gregory}), because cropping B\'ezier patches is simply realized by setting $d(\mathbb{D})$ to zero
and using the inner control points of a B\'ezier patch instead of calculating $p_{i, j}^\textrm{min}(\mathbb{D})$.

\subsection{Iterative Hierarchical Subdivision}
\label{sec:iterative_hierarchical_subdivision}

In order to intersect a ray and a parametric patch, we start with the unit square as the parameter domain, which becomes
repeatedly partitioned, alternating splits along the $u$ and $v$ parameter axes. Instead of an obvious recursive
approach, \Cref{alg:hierarchical_subdivision} traverses the implicit complete binary tree of nested subdomains
iteratively.

Therefore both bounding boxes of the two sub-patches resulting from partitioning are intersected with the ray. Avoiding a
stack, intersection information now is stored in a bit trail \cite{Hughes:2009}, one for each parameter. Only if the ray
intersects both bounding boxes, the bit corresponding to the current level of subdivision is set in the trail of the
current axis to indicate that the second domain also needs to be checked for intersection. Once the bit trail contains
only zeros and the current subtree is pruned, traversal is terminated.

Similar to Dammertz et al.~\cite[Sec.~2.1]{Dammertz:2006}, partitioning can be stopped once the $L_1$-norm of the
bounding box diagonal projected onto the screen is less than half a pixel width for primary rays. As there is no
approximation with a bi-linear surface, flatness criteria are not applicable. After terminating partitioning due to
sufficient precision or missing both bounding boxes, the least significant set bit from both trails determines the next
domain checked for intersection and the next partition axis. Now, \Cref{alg:bounding_gregory} determines the inner
control points and the maximum displacement $d$, and in association with \Cref{alg:crop_bezier} re-computes the current
set $p$ of control points for the subdomain specified by position and size as discussed in the previous sections.

Intersecting the two bounding volumes of the sub-patches instead of intersecting only the bounding volume has two main
advantages: First, it reduces the frequency of backtracking. Backtracking is expensive compared to de Casteljau
subdivision, and can be quite often avoided because one of the two subdomains can immediately be pruned. Second, ordered
traversal is as simple as intersecting the sub domain with the closer bounding volume first. While only adding the
negligible overhead of comparing hit distances, this ordering helps avoiding the subdivision of hidden parts of patches.

Interestingly, we can use the same subdivision and pruning procedure also for Gregory patches: After subdividing the
lower bounding B\'ezier patch, we extend the maximum bound by the maximum displacement determined on the coarser scale.
As the lower bounding B\'ezier patch of the coarser level also bounds the one on the current level and the maximum
displacement is monotonously decreasing, the calculated bounding volumes are conservative.

Our experiments show that the reduction of traversal steps and computation of control points nearly halves traversal
time as compared to intersecting only the bounding volume of the patches restricted to the current domain.

We do not calculate the distance to the bounding box $t_{\textrm{cur}}$ again on termination, as it has already been
determined directly after subdivision. Even if the second domain on the same level was also intersected by the ray, it
cannot be closer as the subdomain whose bounding box is closer is checked first. We could therefore also clear the trail
bit to avoid backtracking to the same level, but these cases are so rare that we could not measure any speedup.

For path tracing, we avoid self-intersection by shifting the origin of the subsequent ray along the normal in the
intersection by the $L_1$ distance of the last bounding volume. While this offset is not minimal, it is numerically
preferable over the diagonal of the bounding box and always avoids self-intersection.

\subsection{Numerical Robustness and Watertightness}
\label{sec:numerical_robustness_and_watertightness}

De Casteljau subdivision is the most robust method for bisection of a B\'ezier patch as it only performs additions and
divisions by two. Intersecting rays with patches using the iterative subdivision method described in
\Cref{sec:iterative_hierarchical_subdivision} only reports false positives; false negatives can only be introduced by
numerical imprecision. In theory, the surface intersected by the algorithm would therefore be completely watertight, and
its approximation error would only be an artificial thickness of the patch. It is important to note that this guarantee
also includes independence from the local subdivision depth, which clearly differentiates the approach from others using
bi-linear approximations after subdivision \cite{Benthin:2007,Tejima:2015}, which either require a fixed subdivision
depth or may introduce cracks due to local deviation of subdivision depth.

In practice, special care needs to be taken for intersection of rays with bounding volumes
\cite{Williams:2005,Ize:2013}. Furthermore, the recomputation of control points after backtracking may be numerically
critical as it may result in control points that differ from the ones calculated by subdivision. An analysis of all
included operations may define an upper bound on the deviation which can be used to extend bounding boxes and guarantee
watertightness. On the other hand, this deviation can be dramatically reduced by anchoring patches in the origin. In our
numerical experiments we measured a deviation as low as \SI{0.0001}{\%}, which suggests that the deviation will most
likely be negligible. For some of our meshes the discrepancy between floating point and double precision subdivision is
about similar.

Intersecting patches in local frames instead of using world space coordinates for all bounding boxes improves
performance significantly by reducing their overlap and therefore lowering the number of parametric domains that cannot
be immediately pruned. While the performance often doubles in our test scenes, the transformation to the local frame is
a source of numerical precision issues that may become visible as gaps between adjacent patches. Extending bounding
boxes at the boundary that fall below a certain size threshold introduces an artificial thickness which closes these
gaps.

\subsection{Efficient Parallelization}
\label{sec:efficient_parallelization}

The apparent parallelization of the method over rays suffers from very high register pressure: Even for a B\'ezier patch
already 48 registers are required to maintain the set of control points. As the vast majority of calculations is
performed component-wise, the number of required registers can be reduced by operating with thread groups of three per
ray, using one thread per component of the three-dimensional vectors. Although this approach results in only 30 active
threads for a SIMD width of 32 and requires several shuffles of values between threads, our experiments report a speedup
of up to \SI{33}{\%}. Besides reducing the register pressure, this parallelization scheme also helps reducing the amount
of control flow and data divergence.

In addition, shared state variables can also be maintained across the thread group, which further reduces register
pressure and allows for further distributing computations across the thread group. We measure another performance gain
of up to \SI{10}{\%} after this optimization.

We also use this parallelization scheme for traversal of the top level hierarchy. While there the number of
component-wise computations is relatively low compared to the time spent for memory access and communication between the
threads of a group, we still measured a clear performance advantage over traversal in only one of the three threads of a
group and also over a two step method that first traverses the hierarchy with one thread per ray and enqueues ray-patch
tuples for a subsequent intersection test. \Cref{lst:thread_groups} shows an example for a parallelization with this
scheme.

Another source of control flow divergence is subdivision, which may split along different parametric directions for
different threads of a warp. We avoid this issue completely by transposing the grid of control points after subdivision
and after recalculation (if required). This way we only need to split along one parametric direction. As we are only
interested in the convex hull of the set of control points, its rotation is irrelevant.

\begin{lstlisting}[
	caption={CUDA example for an arbitrary, unrelated calculation using the parallelization scheme using thread groups of three.}, %
	label=lst:thread_groups,
	captionpos=b,
	float=t,
	morekeywords={hmax3,max3,shuffle},
	%
	columns=fixed,
	basewidth=.5em,
	breaklines=true
]
// -------------- one calculation per thread --------------
vec3f p0 = 3.0f * a + 2.0f * b + vec3f(12.0f);
vec3f p1 = 1.0f * a - 3.0f * b + vec3f(7.0f);
float best = hmax3(p0) + hmax3(p1);

// -------------- in thread groups of three ---------------
// component wise operations look exactly the same
float p0 = 3.0f * a + 2.0f * b + 12.0f;
float p1 = 1.0f * a - 3.0f * b + 7.0f;
// shuffles required for operations on multiple components
float best = max3(shuffle(p0, lane0),
                  shuffle(p0, lane1),
                  shuffle(p0, lane2)) +
             max3(shuffle(p1, lane0),
                  shuffle(p1, lane1),
                  shuffle(p1, lane2));
\end{lstlisting}

\begin{algorithm}[htbp]
	\DontPrintSemicolon
	\caption{\textbf{cropBezierPatch:} Calculate control points of a B\'ezier patch cropped to the parametric domain $\lbrack u_0,
	u_1\rbrack\times\lbrack v_0, v_1\rbrack$.}
	\label{alg:crop_bezier}
	\KwIn{A B\'ezier patch defined by its control points $p$ and the domain $\lbrack u_0, u_1\rbrack\times\lbrack v_0, v_1\rbrack$}
	\KwOut{Control points $q$ of the cropped B\'ezier patch}
	$\Delta u \gets \frac{u_1 - u_0}{3}$\;
	$\Delta v \gets \frac{v_1 - v_0}{3}$\;
	\;
	\tcp{Corners}
	$q_{0, 0} \gets evalBezier(p, u_0, v_0)$\;
	$q_{3, 0} \gets evalBezier(p, u_1, v_0)$\;
	$q_{0, 3} \gets evalBezier(p, u_0, v_1)$\;
	$q_{3, 3} \gets evalBezier(p, u_1, v_1)$\;
	\;
	\tcp{Points on the boundary}
	$q_{1, 0} \gets q_{0, 0} + \Delta u \cdot evalBezierDu(p, u_0, v_0)$\;
	$q_{2, 0} \gets q_{3, 0} - \Delta u \cdot evalBezierDu(p, u_1, v_0)$\;
	$q_{0, 1} \gets q_{0, 0} + \Delta v \cdot evalBezierDv(p, u_0, v_0)$\;
	$q_{3, 1} \gets q_{3, 0} + \Delta v \cdot evalBezierDv(p, u_1, v_0)$\;
	$q_{0, 2} \gets q_{0, 3} - \Delta v \cdot evalBezierDv(p, u_0, v_1)$\;
	$q_{3, 2} \gets q_{3, 3} - \Delta v \cdot evalBezierDv(p, u_1, v_1)$\;
	$q_{1, 3} \gets q_{0, 3} + \Delta u \cdot evalBezierDu(p, u_0, v_1)$\;
	$q_{2, 3} \gets q_{3, 3} - \Delta u \cdot evalBezierDu(p, u_1, v_1)$\;
	\;
	\tcp{Inner control points}

	\settowidth{\maxwidth}{$q_{1, 1} \gets$}
	\algalign{q_{1, 1} \gets}{q_{1, 0} + \Delta v \cdot evalBezierDv(p, u_0, v_0)}\;
	\algalign{}{\phantom{q_{1, 0} }+ \Delta u \cdot \Delta v \cdot evalBezierDuDv(p, u_0, v_0)}\;

	\settowidth{\maxwidth}{$q_{2, 1} \gets$}
	\algalign{q_{2, 1} \gets}{q_{3, 1} - \Delta u \cdot evalBezierDu(p, u_1, v_0)}\;
	\algalign{}{\phantom{q_{3, 1} }- \Delta u \cdot \Delta v \cdot evalBezierDuDv(p, u_1, v_0)}\;

	\settowidth{\maxwidth}{$q_{1, 2} \gets$}
	\algalign{q_{1, 2} \gets}{q_{1, 3} - \Delta v \cdot evalBezierDv(p, u_0, v_1)}\;
	\algalign{}{\phantom{q_{1, 3} }- \Delta u \cdot \Delta v \cdot evalBezierDuDv(p, u_0, v_1)}\;

	\settowidth{\maxwidth}{$q_{2, 2} \gets$}
	\algalign{q_{2, 2} \gets}{q_{2, 3} - \Delta v \cdot evalBezierDv(p, u_1, v_1)}\;
	\algalign{}{\phantom{q_{2, 3} }+ \Delta u \cdot \Delta v \cdot evalBezierDuDv(p, u_1, v_1)}\;
	\Return{$q$}
\end{algorithm}

\begin{algorithm}
	\DontPrintSemicolon
	\caption{\textbf{calcPointsAndD:} Calculating control points of a B\'ezier patch as a lower bound and the maximum displacement of an upper
	bounding B\'ezier patch of a Gregory Patch restricted to a domain. For a B\'ezier patch the lower bounding patch
	equals the patch itself and the displacement of the upper bound is zero.}
	\label{alg:bounding_gregory}
	\small
  \KwIn{A B\'ezier/Gregory patch with control points $p$ and the domain to which it is restricted $[u_0, u_1]\times[v_0, v_1]$ and a flag $isGregoryPatch$}
  \KwOut{A set of control points $q$ of the lower bounding B\'ezier patch and the maximum distance $d$ of the upper bounding B\'ezier patch}
			\uIf{isGregoryPatch}{
      $p_{1,1}^{min} \gets \frac{u_0 \cdot p_{1,1}^u + v_1 \cdot p_{1,1}^v}{u_0 + v_1}$\;
      $p_{1,1}^{max} \gets \frac{u_1 \cdot p_{1,1}^u + v_0 \cdot p_{1,1}^v}{u_1 + v_0}$\;

      $p_{2,1}^{min} \gets \frac{(1 - u_0) \cdot p_{2,1}^u + v_0 \cdot p_{2,1}^v}{(1 - u_0) + v_0}$\;
      $p_{2,1}^{max} \gets \frac{(1 - u_1) \cdot p_{2,1}^u + v_1 \cdot p_{2,1}^v}{(1 - u_1) + v_1}$\;

      $p_{1,2}^{min} \gets \frac{u_1 \cdot p_{1,2}^u + (1 - v_1) \cdot p_{1,2}^v}{u_1 + (1 - v_1)}$\;
      $p_{1,2}^{max} \gets \frac{u_0 \cdot p_{1,2}^u + (1 - v_0) \cdot p_{1,2}^v}{u_0 + (1 - v_0)}$\;

      $p_{2,2}^{min} \gets \frac{(1 - u_1) \cdot p_{2,2}^u + (1 - v_0) \cdot p_{2,2}^v}{(1 - u_1) + (1 - v_0)}$\;
      $p_{2,2}^{max} \gets \frac{(1 - u_0) \cdot p_{2,2}^u + (1 - v_1) \cdot p_{2,2}^v}{(1 - u_0) + (1 - v_1)}$\;

      $w_{u, 1} \gets
				\begin{cases}
					\tfrac{1}{3} & u_0 \leq \tfrac{1}{3} \land u_1 \geq \tfrac{1}{3},\\
					u_1          & u_1 \leq \tfrac{1}{3},\\
					u_0          & \textrm{else}.
				\end{cases}$\;

      $w_{u, 2} \gets
			\begin{cases}
				\tfrac{2}{3} & u_0 \leq \tfrac{2}{3} \land u_1 \geq \tfrac{2}{3},\\
				u_1          & u_1 < \tfrac{2}{3},\\
				u_0          & \textrm{else}.
			\end{cases}$\;

      $w_{v, 1} \gets
			\begin{cases}
				\tfrac{1}{3} & v_0 \leq \tfrac{1}{3} \land v_1 \geq \tfrac{1}{3},\\
				v_1          & v_1 < \tfrac{1}{3},\\
				v_0          & \textrm{else}.
			\end{cases}$\;

      $w_{v, 2} \gets
			\begin{cases}
				\tfrac{2}{3} & v_0 \leq \tfrac{2}{3} \land v_1 \geq \tfrac{2}{3},\\
				v_1          & v_1 < \tfrac{2}{3},\\
				v_0          & \textrm{else}.
			\end{cases}$\;

      $w_{u, 1}^{max} \gets 3 \cdot w_{u, 1} \cdot (1 - w_{u, 1}) \cdot (1 - w_{u, 1})$\;
      $w_{u, 2}^{max} \gets 3 \cdot w_{u, 2} \cdot w_{u, 2} \cdot (1 - w_{u, 2})$\;
      $w_{v, 1}^{max} \gets 3 \cdot w_{v, 1} \cdot (1 - w_{v, 1}) \cdot (1 - w_{v, 1})$\;
      $w_{v, 2}^{max} \gets 3 \cdot w_{v, 2} \cdot w_{v, 2} \cdot (1 - w_{v, 2})$\;

      $p_{1,1} \gets min\{p_{1,1}^{min}, p_{1,1}^{max}\}$\;
      $p_{2,1} \gets min\{p_{2,1}^{min}, p_{2,1}^{max}\}$\;
      $p_{1,2} \gets min\{p_{1,2}^{min}, p_{1,2}^{max}\}$\;
      $p_{2,2} \gets min\{p_{2,2}^{min}, p_{2,2}^{max}\}$\;

			\settowidth{\maxwidth}{$d \gets\ $}%
			\algalign{d \gets }{\phantom{+\ \ }w_{u, 1}^{max} \cdot w_{v, 1}^{max} \cdot |p_{1,1}^{max} - p_{1,1}^{min}|}\;
			\algalign{}{+w_{u, 2}^{max} \cdot w_{v, 1}^{max} \cdot |p_{2,1}^{max} - p_{2,1}^{min}|}\;
			\algalign{}{+w_{u, 1}^{max} \cdot w_{v, 2}^{max} \cdot |p_{1,2}^{max} - p_{1,2}^{min}|}\;
			\algalign{}{+w_{u, 2}^{max} \cdot w_{v, 2}^{max} \cdot |p_{2,2}^{max} - p_{2,2}^{min}|}\;
			}
			\Else {
			$d \gets 0$
			}
			$q \gets\ $cropBezierPatch($p, u_0, u_1, v_0, v_1$)\;

			\Return{$(q, d)$}
\end{algorithm}

\begin{algorithm}
	\DontPrintSemicolon
  \caption{Intersecting a B\'ezier or Gregory patch.}
	\label{alg:hierarchical_subdivision}
	\small
  \KwIn{A $ray$ with its current closest hit distance $t_{\textrm{max}}$, the patch control points $p_{\textrm{org}}$ and a flag $isGregoryPatch$}
  \KwOut{Intersection details and distance $t_{\textrm{max}}$}
  $pos \gets$ (0, 0)\;
  $size \gets$ ($2^{23}$, $2^{23}$)\;
  $trail \gets$ (0, 0)\;
  $axis \gets$ 0\;
	$t_{\textrm{max}} \gets ray.t_{\textrm{max}}$\;
  \uIf{isGregoryPatch} {
    $(p, d) \gets $calcPointsAndD($isGregoryPatch, p_{\textrm{org}}, pos, size$)\;
  }
  \Else {
    $p \gets p_{\textrm{org}}$\;
    $d \gets (0, 0, 0)^T$\;
  }
  \While{true}{
    $subdividing \gets$ \textbf{false}\;
    \uIf{$\lnot$terminate($p, ray, pos, size$)} {
      $(p_L, p_R) \gets$ subdivideDeCasteljau($p, axis$)\;
      ($hitL, hitR, closest, t_{\textrm{cur}}$) $\gets$ intersect($ray, t, p_L, p_R, d$)\;
      \If{$hitL$ \textbf{or} $hitR$} {
        $subdividing \gets$ \textbf{true}\;
        $size[axis] \gets size[axis] / 2$\;
        \If{$hitL$ \textbf{and} $hitR$}{
          $trail[axis] \gets trail[axis] \oplus size[axis]$\;
        }
        \uIf{$closest = R$ \textbf{or} $\lnot hitL$} {
          $p \gets p_R$\;
          $pos[axis] \gets pos[axis] \oplus size[axis]$\;
        }
        \Else {
          $p \gets p_L$\;
        }
        $axis \gets 1 - axis$\;
      }
    }
    \Else {
      \If{$t_{\textrm{cur}} < t_{\textrm{max}}$} {
        $t_{\textrm{max}} = t_{\textrm{cur}}$\;
        $intersection \gets$ $(pos, size, ...)$\;
      }
    }
    \If{$\lnot subdividing$}{
      \lIf{$trail = (0, 0)$} {\textbf{break}}
      $lvl \gets$ countTrailingZeros($trail$)\;
      $axis \gets$ argmin($lvl$)\;
      \lIf{$axis = 0$}{
        $size \gets (2^{lvl[0]}, 2^{lvl[0]+1})$
      }
      \lElse{
        $size \gets (2^{lvl[1]}, 2^{lvl[1]})$
      }
      $pos[axis] \gets pos[axis] \oplus size$\; %
      $trail[axis] \gets trail[axis] \oplus size$\;
      $pos \gets pos \land \lnot(size - 1)$\; %
      $axis \gets 1 - axis$\;
    }
    \If{$\lnot subdividing$ \textbf{or} $isGregoryPatch$}{
      $(p, d) \gets $calcPointsAndD($isGregoryPatch, p_{\textrm{org}}, pos, size$)
    }
  }
  \Return{$(t_{\textrm{max}}, intersection)$}
\end{algorithm}

\section{Results and Discussion}

\begin{figure*}
	\centering
	\begin{tabular}{@{}ccc@{}}
		\includegraphics[width=.32\textwidth]{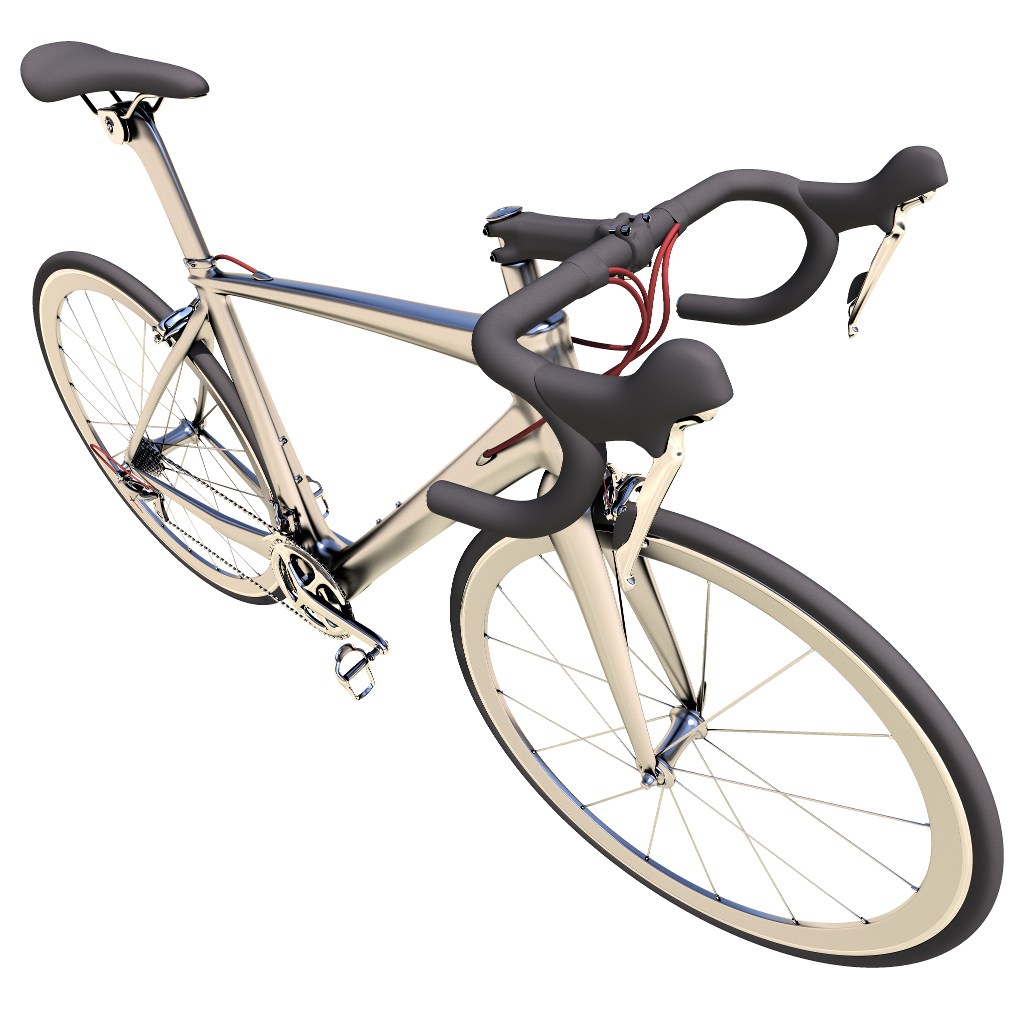}&%
		\hspace*{.15\textwidth}&
		\includegraphics[width=.32\textwidth]{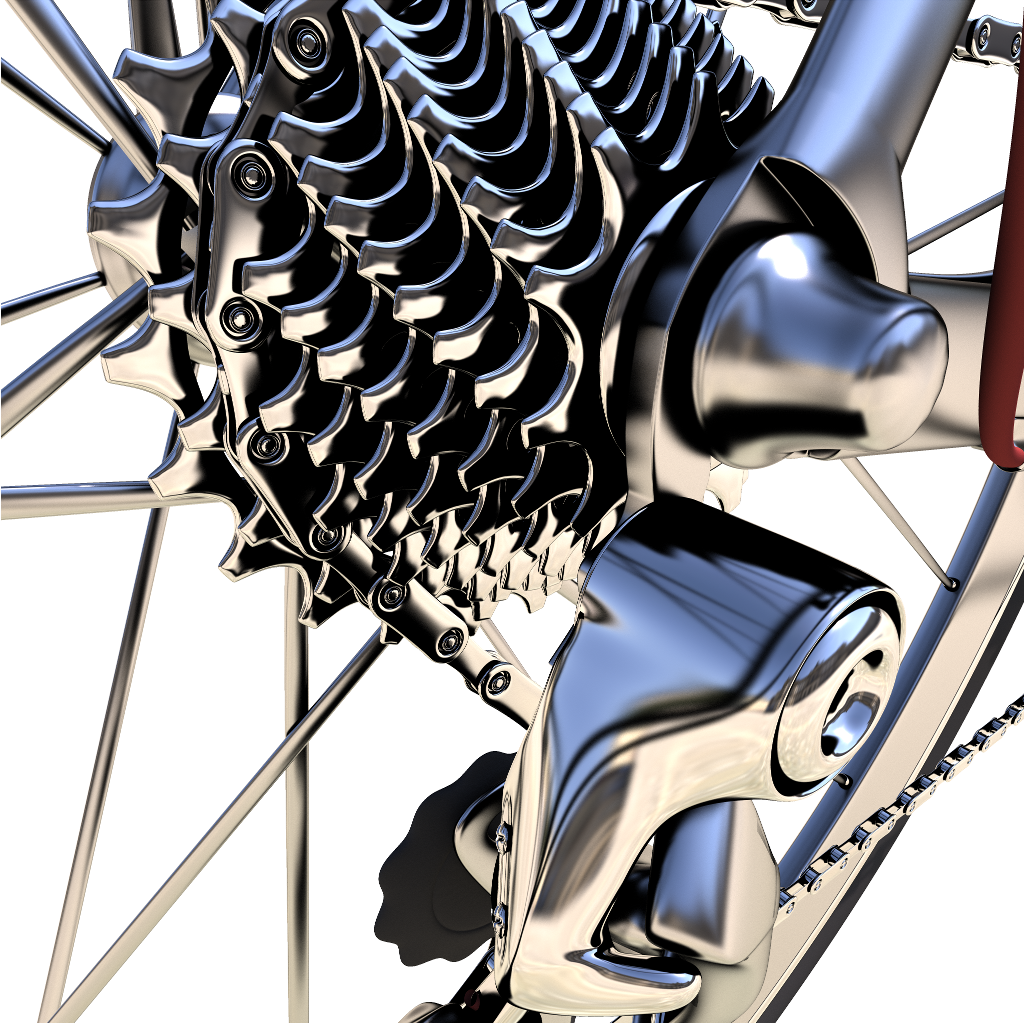}\\%
		{\small \SI{185}{\million rays/s} (\SI{60}{\million rays/s})} && {\small \SI{73}{\million rays/s} (\SI{51}{\million rays/s})}\\%
		\multicolumn{3}{c}{{\small a) Road Bike (courtesy of Yasutoshi ``Mirage'' Mori)}}\\%
		\vspace*{1em}\\%
		\includegraphics[width=.32\textwidth]{ArmorGuyRun11-Textured2-postprocessed.png}&%
		\hspace*{.15\textwidth}&
		\includegraphics[width=.32\textwidth]{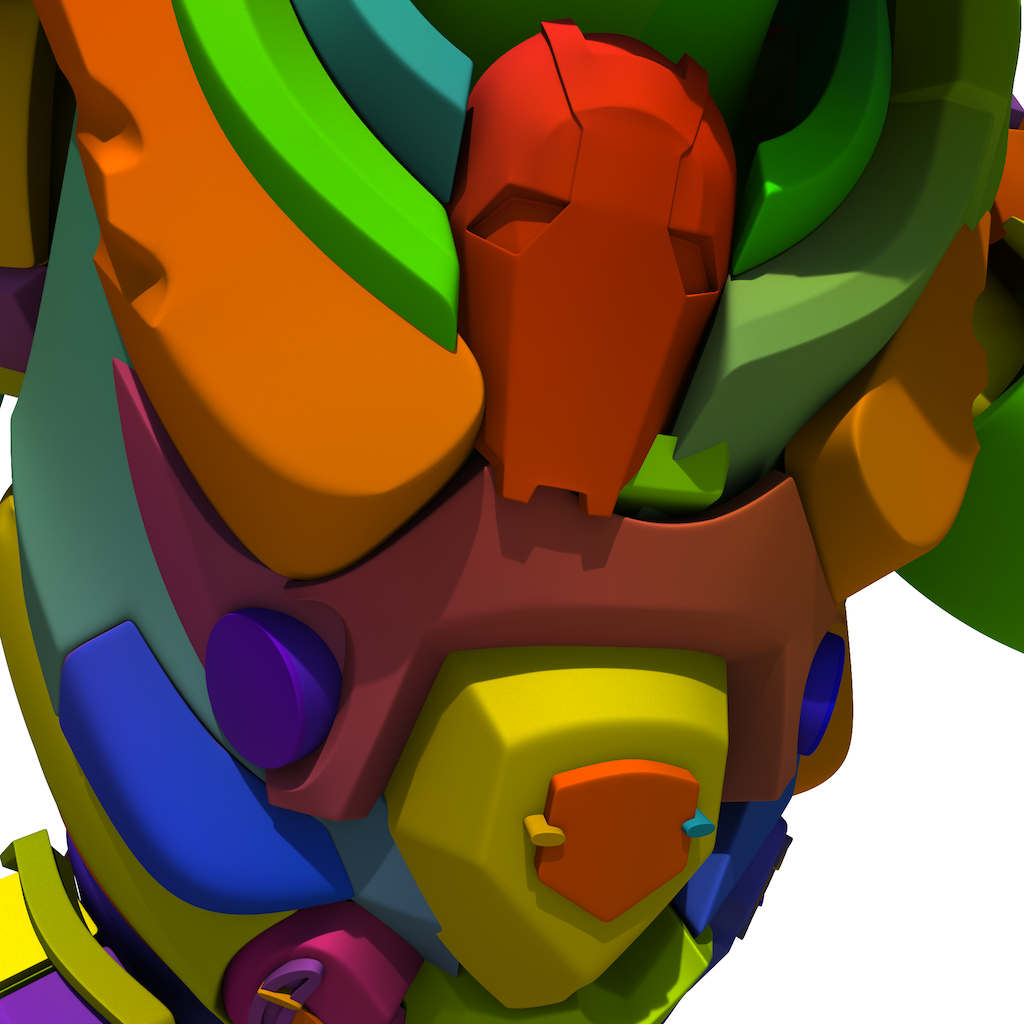}\\%
		{\small\SI{220}{\million rays/s} (\SI{85}{\million rays/s})} && {\small\SI{65}{\million rays/s} (\SI{62}{\million rays/s})}\\%
		\multicolumn{3}{c}{{\small b) Armor Guy and false color closeup of texture coordinates (courtesy of DigitalFish)}}\\%
		\vspace*{1em}\\%
		\includegraphics[width=.32\textwidth]{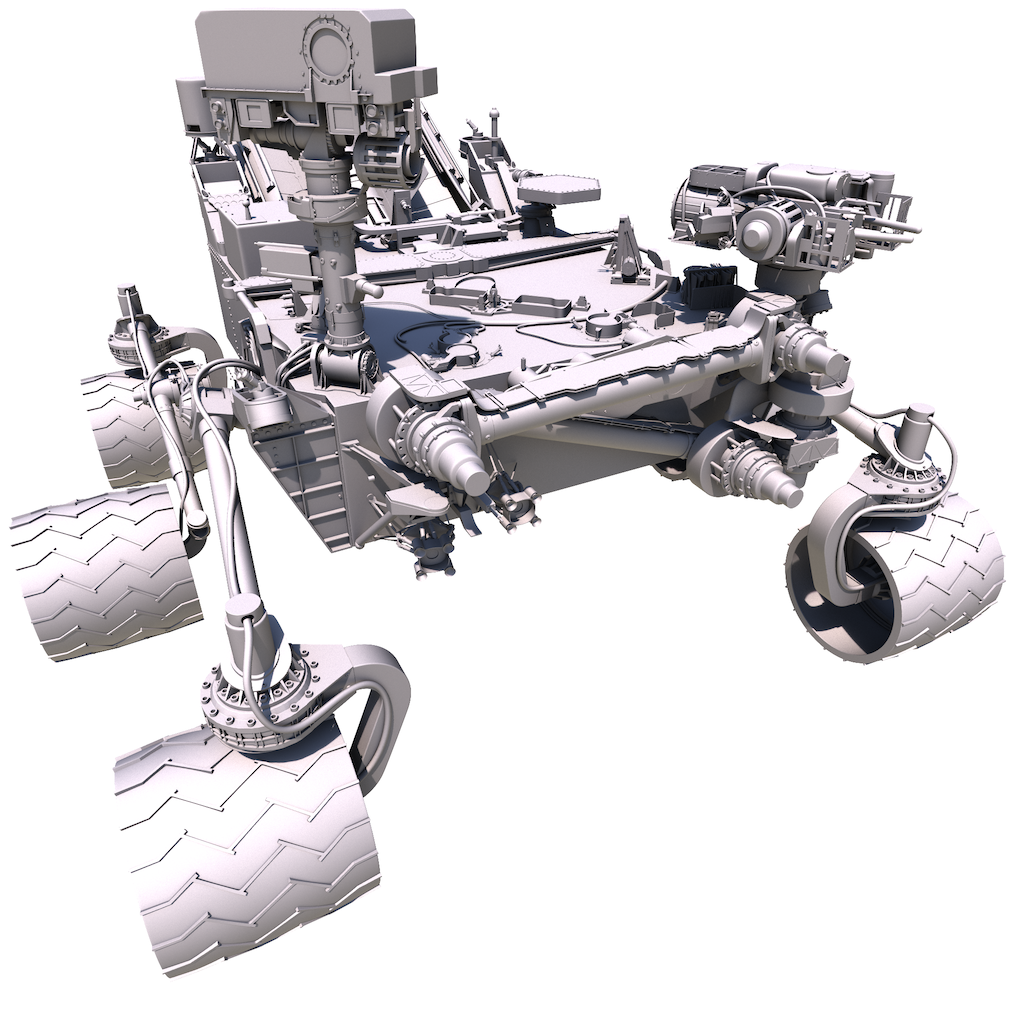}&%
		\hspace*{.15\textwidth}&
		\includegraphics[width=.32\textwidth]{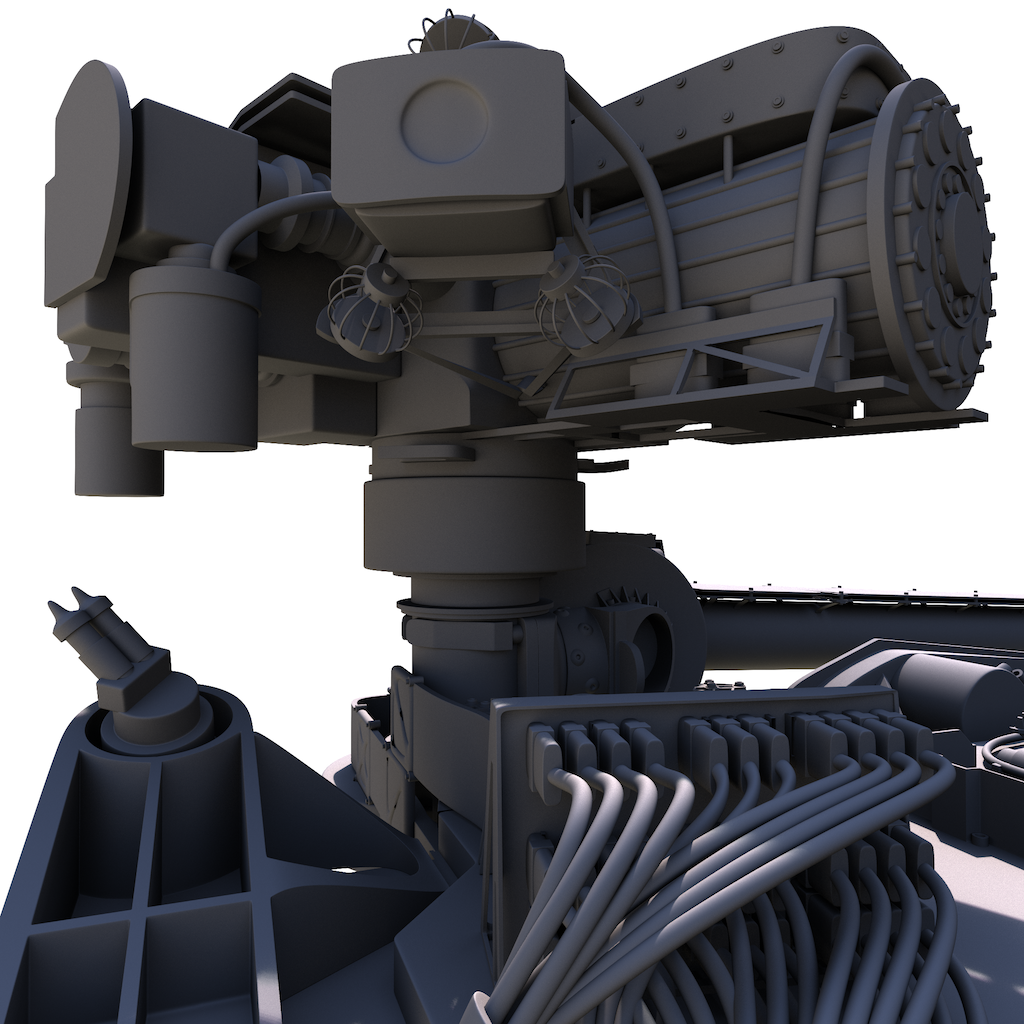}\\%
		{\small \SI{131}{\million rays/s} (\SI{72}{\million rays/s})} && {\small\SI{70}{\million rays/s} (\SI{43}{\million rays/s})}\\%
		\multicolumn{3}{c}{{\small c) Curiosity (courtesy of DigitalFish)}}
	\end{tabular}
	\caption{Several meshes, ray traced using the described method and the performance for primary (diffuse) rays measured on a NVIDIA Titan V\texttrademark\ GPU.}
	\label{fig:results}
\end{figure*}

We evaluated the performance of our method using a path tracer with next event estimation on several different
generations of NVIDIA GPUs. The numbers presented in this article were obtained using a NVIDIA Titan V\texttrademark.
Due to the simplicity of the models only one primary, one reflection, and one shadow ray to the light sources were
traced.

Employing a basic wavefront ray tracer \cite{Laine:2013}, we can distinguish intersection performance from ray setup and
shading. We do not perform compaction of the ray stream after intersecting primary rays; intersection performance for
secondary rays is therefore not only penalized by increased incoherency of secondary rays, but also by the resulting
lower SIMD occupancy.

Intersection performance for larger models is mostly dominated by traversal performance of the hierarchy referencing
the patches. We use a simple binary Bounding Volume Hierarchy (BVH), recursively constructed using bins to split
approximately according to the Surface Area Heuristic (SAH) \cite{Goldsmith:1987,Havran:2000,Shevtsov:2007,Wald:2007}.
Traversal performance can potentially be improved by utilizing spatial splits \cite{Stich:2009} also for patches.
Furthermore, while we traverse a binary hierarchy using a node index stack and persistent threads \cite{Aila:2009},
recent advances in stackless BVH traversal \cite{Binder:2016} and fast traversal of wide BVHs on GPUs \cite{Ylitie:2017}
will certainly also improve the overall performance.

The results in the captions of \Cref{fig:teaser,fig:results} show that the direct ray tracing of the bicubic patches
resulting from feature adaptive subdivision can be performed interactively or even in real time. Depending on the scene
complexity, the pure intersection performance of pre-tessellated meshes outperforms the presented direct intersection
clearly, whereas memory limitations easily can become prohibitive depending on the desired precision. However, taking
into account the time required for tessellation and the increased time for bounding volume hierarchy construction,
direct intersection in many cases is faster and thus favorable.

\section{What about Displacements?}

Applying displacements to patches complicates the determination of the
bounding boxes, as now bounds on both the displacement as well as the range of directions along which the patch is
displaced are required.

Bounds on the displacement can be either scalar or vector valued. Efficient lookup requires a hierarchical structure
such that the current displacement can be determined with a minimal amount of read
operations \cite{Williams:1983}. Displacement maps can be given per patch \cite{Burley:2008} or in one or multiple textures, which are then
accessed using texture coordinates. For the latter bounds cannot be read from a single location in this texture due to
a potential warping of the domain. As the size of the hierarchy built for vector valued displacement textures equals the
size required for a bounding volume hierarchy of a pre-subdivided patch, using direct intersection with displacements
only saves memory if the displacement can be used multiple times, e.g. for motion blur. Depending on their complexity,
bounds for procedural displacement of the patch restricted to a parametric domain may be determined quite efficiently
\cite{Heidrich:1998,Velazquez-Armendariz:2009}.

While bounds of the possible range of normal directions can be computed using normal cones or tangent cones
\cite{Munkberg:2010}, it adds yet another significant overhead.

Displacing a tessellated subdivision surface may remove all mathematical guarantees
of the underlying surface. Often, the level of tessellation is selected as level
of detail \cite{pixarrt06}. While practical in the REYES architecture and rasterization, the visibility
may be affected and shadows and silhouettes may be changed \cite{Dammertz:2006}, for example.

The ensemble of these issues raises the interesting question whether a subdivided mesh with a displacement, as used for modeling, is a representation suitable for efficient rendering at all. We therefore see our algorithm
as an initial step to start exploring displacements that allow for level of detail
and at the same time provide mathematical guarantees for efficient evaluation and
visibility error control in future research.

\section{Conclusion}

We have presented a numerically robust algorithm that intersects rays with B\'ezier and Gregory patches with very high
performance and precision on massively parallel hardware based on an improved calculation of bounds of Gregory patches,
iterative hierarchical subdivision, and a special parallelization scheme. After creating these patches
using feature adaptive
subdivision of Catmull-Clark subdivision meshes, they can be efficiently and directly intersected with
rays.

Due to the superiority of hierarchical culling over brute force (hardware) tessellation and rasterization, direct ray
tracing of these meshes can in some cases even outperform rasterization, which is expected especially for large scenes
with a high depth complexity and very high precision requiring fine tessellation for rasterization.

While one might argue about the quality of the approximation of irregular patches by Gregory patches, consistent and
thus predictable output from modeling until final rendering may be more important. Hence our new algorithm is
beneficial during modeling and animation playback, especially when caching and tessellation are inefficient
and even more so for the large number of applications that do not require displacement.

\newcommand{\etalchar}[1]{$^{#1}$}

\end{document}